\documentclass{article}
\usepackage{anyfontsize}
\usepackage[most]{tcolorbox}

\usepackage[utf8]{inputenc}
\usepackage[T1]{fontenc} 
\usepackage[russian, english]{babel}

\usepackage[preprint]{neurips_2025}

\usepackage[utf8]{inputenc} 
\usepackage[T1]{fontenc}    
\usepackage{hyperref}       
\usepackage{url}            
\usepackage{booktabs}       
\usepackage{amsfonts}       
\usepackage{nicefrac}       
\usepackage{microtype}      
\usepackage{xcolor}         
\usepackage{graphicx}
\usepackage{wrapfig}
\usepackage{amsmath}
\usepackage{bbm}
\usepackage{enumitem}
\usepackage{makecell}  
\usepackage{colortbl}
\usepackage{pifont}

\usepackage{amsmath,amssymb,graphicx,hyperref,algorithm,algorithmicx,algpseudocode}
\usepackage{geometry}

\definecolor{mygray}{gray}{.94}

\title{Developing a Grounded View of AI}

\author{%
Bifei Mao, Lanqing Hong\\
Huawei Noah's Ark Lab
}

\begin{document}

\maketitle


\begin{abstract}  

As a capability coming from computation, how does AI differ fundamentally from the capabilities delivered by rule-based software program? The paper examines the behavior of artificial intelligence (AI) from engineering points of view to clarify its nature and limits. The paper argues that the rationality underlying humanity's impulse to pursue, articulate, and adhere to rules deserves to be valued and preserved. Identifying where rule-based practical rationality ends is the beginning of making it aware until action. Although the rules of AI behaviors are still hidden or only weakly observable, the paper has proposed a methodology to make a sense of discrimination possible and practical  to identify the distinctions of the behavior of AI models  with three types of decisions. It is a prerequisite for human responsibilities with alternative possibilities, considering how and when to use AI. It would be a solid start for people to ensure AI system soundness for the well-being of humans, society, and the environment. 

\end{abstract}  

\section{Introduction}

Developing a sound conception of AI is essential for deciding when and how to adopt it. There are various products and services on the market, each of which adopts different technologies. When a product is declared AI inside, AI is one of these technologies. It is very important to be aware of its limitations while accepting its advantages~\citep{jensen2024reflection,johnson2025sociotechnical,watts2017should}.

There is enthusiastic imagination about AI, even indescribable AGI. The rational question raised from material cause is: as a capability coming from computation, how does AI differ fundamentally from the capabilities delivered by rule-based software program?

While researchers and promoters of AI frequently use metaphorical language, a lot of ambiguity and conceptual vagueness are used. Recent work on AI metaphors and technological promises shows how such rhetoric can both clarify and obscure what AI systems actually do, often relying on fuzzy conceptual metaphors and myths of intelligent machines. Avoiding ambiguity is quite a basic and general mindset of engineering. There is ample room for investigation.

 For easy discussion, we introduce the notation of terminology and parties:

\begin{tcolorbox}[title=Notation and Terminology \#1, colback=gray!5, colframe=black!60, coltitle=black, boxrule=0.6pt]

\textbf{Program \foreignlanguage{russian}{Я}}: Software that embodies artificial intelligence functionality, capable of natural language interface, autonomous reasoning and adaptive behavior. 

\textbf{Program $\Phi$}: Software that realizes a given function by strictly following the rules inherent in the use. 

\end{tcolorbox}

The Cyrillic letter \foreignlanguage{russian}{Я}, pronounced “ya” (“I” in English), symbolizes the self-referential nature of AI that imitates human cognition. Its reversed form visually reinforces this idea of mirroring human intelligence. The Greek letter $\Phi$—symmetrical and balanced—represents equilibrium between creativity and adherence to rules.  The terms “pictograph” and “ideograph” used here refer to two of the Six Principles of Chinese character formation, illustrating how visual forms can convey both shape and meaning.

Building upon the above definitions, we further distinguish among the key parties involved in the development, deployment, and use of AI systems. These representative roles help clarify the perspectives and responsibilities referenced throughout the subsequent analysis.

\begin{table}[h!]
\centering
\caption{Notation for Parties}
\begin{tabular}{ll}
\toprule
\textbf{Notation} & \textbf{Roles} \\
\midrule
David & AI researcher and programmer \\
Bob & System engineer of AI product; also serves as design engineer \\
Jack & AI user, sometimes as deploy engineer \\
\bottomrule
\end{tabular}
\end{table}

The problem domain and the solution domain represent related yet distinct perspectives in engineering point of view. Similarity with behavioral analysis and functional analysis. This paper will explore the question as mentioned above by considering AI as a scientific discovery, an engineering innovation artifact, and a human-made phenomenon. After that, it carries about a concise terminology and epistemic framework with high semantic entropy, which will be useful and practical for Jack and Bob becoming prudence on when and how to use AI.

\section{View AI as a Scientific Discovery?}

The field of AI has generated an extraordinary volume of research within a thriving scientific ecosystem. Can AI capabilities be considered as scientific discoveries?

The universal approximation theorem~\citep{cybenko1989approximation,hornik1989multilayer} demonstrates that a feedforward neural network with a single hidden layer can approximate any continuous function to arbitrary precision under certain conditions. Yet this theorem belongs to the solution domain: it reveals the existence of approximation capability but does not specify which functions in the actual problem domain are being approximated.

Similarly, empirical scaling laws summarize the relationships among model parameters, data size, and computational resources~\citep{hoffmann2022training,kaplan2020scaling}. These relations capture performance and sustainability trends from the solution perspective, but they do not directly explain the correspondence between AI behavior and the underlying real-world problems it aims to solve.

From the viewpoint of its behavior, the leaderboard scores achieved by Program~\foreignlanguage{russian}{Я} serve as indirect evidence of performance. However, the pre-defined test questions and answers differ from the open-ended inputs and outputs encountered in real-world applications. A generative model produces an output $y$ sampled from a conditional distribution $p(y|x)$, where $x$ may represent text, images, or their combinations (e.g., a user asking, “How can I fix a broken lamp?”). The model’s response, such as “To fix a broken light, you need~[...],” reflects a probabilistic mapping between symbolic sequences $x = (w_{x1}, \dots, w_{xk})$ and $y = (w_{y1}, \dots, w_{ys})$. Yet this symbolic correspondence captures only part of the true relationship in the problem domain—the semantic mapping between meanings $s_x$ and $s_y$ must still be inferred and completed by humans themselves~\citep{bender2021dangers}.

While AI models such as program \foreignlanguage{russian}{Я} are often celebrated  as  scientific achievements, one may expect a general law describing  the relationship between its inputs and outputs, typically expressed as $y = f(x, \xi)$. However, the concrete law governing the behavior of such program \foreignlanguage{russian}{Я} in their real problem domain remains largely concealed.

\section{View AI as a Technical Artifact of Engineering Innovation?}

This raises a related question: what if program \foreignlanguage{russian}{Я}  is not viewed as a scientific discovery, but as a technical artifact of engineering innovation?

Human engineering and technological practices have a venerable history. At its core, engineering addresses human needs, guided by the paramount purpose of serving humanity. It exhibits inherent reproducibility in its technical artifacts and methods, including the associated human activities. In general, an engineered artifact, i.e., program $\Phi$, is designed by the engineer Bob in accordance with selected rules to achieve given functions and performance goals, and created to meet its design specification as per appropriate quality assurance measures, so that its behavior can be predicated according to preset rules.

Behind these practices lies the accumulated body of practical and integrated engineering practices developed over years by generations of engineers like Bob, which act as rules and guides for new generation engineer Bob to deeply understand the essence of the system, to design systems that meet designated performance objectives, and to eliminate as far as possible the interactions that may lead to system failures.

To illustrate the difference about AI, consider the following examples.

\begin{tcolorbox}[title=Fact as Example \#1, colback=gray!5, colframe=black!60, coltitle=black, boxrule=0.6pt]

On May 1, 2023, Geoffrey Hinton announced his resignation from Google in an interview with The New York Times. The following day, in an interview with MIT Technology Review , he had an answer for confabulations of AI, bullshitting is a feature, not a bug~\citep{heaven2023scared,kovtun2023geolocation,nosta2023hallucinations}. When Hinton made this remark, he was pointing to the fact, making stuff up isn’t the problem, he had once also expected computers to be either right or wrong—not something in between.

\end{tcolorbox}

\begin{tcolorbox}[title=Fact as Example \#2, colback=gray!5, colframe=black!60, coltitle=black, boxrule=0.6pt]

On May 20, 2025, at the Sequoia AI Summit in Los Angeles, the third moderator, Konstantine Buhler introduced the keyword stochastic mindset~\citep{buhler2025stochastic}. "This is a departure from traditional deterministic thinking, born out of traditional software development where you program a system to do a thing and it will always do that thing. Given a set of inputs A, you will always get B. We love that. It’s so comforting to live in this world. It’s binary. There’s not grey."

"Now, we're entering an era of computing that's going to be stochastic. If you ask a computer to remember the number 73, it'll remember that tomorrow, next week, next month. If you ask a person or an AI, well, it might remember 73. It might remember 37, 72, 74, the next prime, 79, or nothing at all. The unpredictable nature is a feature, not just a bug." "This shift will require everyone to embrace 'way more leverage with significantly less certainty'”

\end{tcolorbox}

\begin{tcolorbox}[title=Fact as Example \#3, colback=gray!5, colframe=black!60, coltitle=black, boxrule=0.6pt]
320 BC, Mencius explains:  
Being asked to carry Mount Tai to the North Sea and saying to others “I cannot” is genuinely a matter of inability. 

Being asked to break a branch for an elder and saying “I cannot” is not inability but unwillingness~\citep{ivanhoe2005readings}.
\end{tcolorbox}

Geoffrey Hinton was among the first prominent AI scientists to publicly warn the risks of AI~\citep{bryant2023chatbots,heaven2023scared,metz2023godfather}, while simultaneously pointing out that errors in generative models—known as ``hallucinations'' by AI researchers and often seen as a fatal flaw in the technology—are not a bug of program \foreignlanguage{russian}{Я} but a feature of its design. Similarly, its unpredictable nature. One may agree or disagree with these statements. Yet remarks of this earth-shattering nature, in essence, amount to David's declaration of ``I cannot''—not that he would not, but that he could not.

More than the foregoing, they mark a fundamental departure from traditional computer engineering. The contention that deterministic thinking arises from program $\Phi$ doing a thing and it will always do that thing embodies a profound conflation itself. In reality, such behavior merely means computational repeatability and functional high quality. And for the software functionality which program $\Phi$ addresses, the design principle may be either deterministic thinking or probabilistic thinking, which depends on whether the real-world characteristics are inherently deterministic or probabilistic.

Beyond design, quality assurance is therefore an indispensable engineering practice—the foundation from which reliability and trust emerge. Here, quality does not mean “good,” “luxurious,” or “shiny” in the everyday sense, but rather to the degree to which a product conforms to its specified requirements. To say that a product is of high quality is equivalent to say that its actual performance consistently meets its expected function. 

More precisely, this idea of conformance is often treated probabilistically. Suppose a system can be expressed as $y = f(x, \xi)$; then for any given input $x$, the difference between actual $y$ and expected $y$ can be viewed as a random variable whose variations follow regular statistical laws such as the law of large numbers and the central limit theorem. These assumptions support traditional engineering artifacts, where uncertainty can be bounded and quality can be statistically verified.

Program~$\Phi$, as a conventional engineered system, is expected to behave deterministically—it is either correct or incorrect, with no ambiguity in between. Program~\foreignlanguage{russian}{Я}, however, departs fundamentally from this paradigm. Its behavior cannot be fully specified in advance, nor can its outputs be consistently compared with a single expected result. The notion of “expected output” $y_{\text{expect}}(x)$ is not able to be predefined for most real-world inputs, rendering probabilistic assumptions about deviation, such as a normal distribution of $(y_{\text{actual}} - y_{\text{expect}})$, unjustified.

This leads to a deeper question: how could we understand Program~\foreignlanguage{russian}{Я} when it can be regarded as neither a conventional scientific discovery nor a traditional engineered artifact, besides myth or metaphor? 

\section{View AI as a Philosophical Object}

Taking one step further to the ground, we now only consider behaviors of program \foreignlanguage{russian}{Я} as human-made stochastic phenomenon. 

It has long been recognized that when the process of a natural or engineered event is sufficiently stable, it can often be modeled by a probability distribution, such as the normal, $t$, $F$, Beta, or Gamma distribution~\citep{box1976science}. For instance, annual variations in temperature or precipitation have been successfully treated using stochastic models that reproduce their empirical probability distributions and temporal dependence, showing how abstract probabilistic models can yield remarkably accurate forecasts in hydroclimatic settings~\citep{parlange2000extended}. This predictive success has sometimes led to the overextension of probabilistic reasoning into domains where stability and regularity are not guaranteed, as highlighted by critiques of applying thin-tailed statistical models in environments dominated by rare, high-impact events and by warnings that not all uncertainties can be meaningfully expressed as probabilities.

Yet, what are the stable probabilistic fluctuations underlying the behavior of Program~\foreignlanguage{russian}{Я}? At present, these remain unknown. The state of “unknown,” however, is itself meaningful—it indicates that something essential remains concealed and open to investigate~\citep{ji2023imbalanced}. 

As the limits of language delineate the limits of thought, only by naming can one begin to observe the boundaries of the unknown.
We have distinguished two kinds of relationships that characterize the inherent attributes of the objects to be understand:

\begin{tcolorbox}[title=Notation and Terminology \#2, colback=gray!5, colframe=black!60, coltitle=black, boxrule=0.6pt]

\textbf{\foreignlanguage{russian}{Я}-Relation}: 
A stochastic or partially indeterminate relationship, in which rules are hidden or only weakly observable—especially in the behaviors of Program \foreignlanguage{russian}{Я} as a human-made phenomenon.

\textbf{$\Phi$-Relation}: A relationship inheres within, or is subject to, nature and social laws and rules. This term encompasses the rule-based rational outcomes in domains such as science, engineering, law, medicine, economics, and politics and others. 

\end{tcolorbox}

Correspondingly, we introduce two modes of reasoning:

\begin{tcolorbox}[title=Notation and Terminology \#3, colback=gray!5, colframe=black!60, coltitle=black, boxrule=0.6pt]

\textbf{\foreignlanguage{russian}{Я}-Thinking}: The cognition and action behavior of Program \foreignlanguage{russian}{Я} that mimics or diverges from human activities.

\textbf{$\Phi$-Thinking}: The cognition and action behavior to pursue, articulate, and adhere to rules,  in which humans master rules, transcend them, and then return to an enriched set of rules. 

\end{tcolorbox}

$\Phi$-Thinking constitutes a subset of human cognitive activities. The complementary spectrum – encompassing intuition, imagination, artistic expression, emotional instincts, superstition, logical fallacies, and lapses or pauses in critical thinking, among others – is collectively termed non-$\Phi$-thinking. Both $\Phi$-Thinking and non-$\Phi$-thinking become potential targets for AI to simulate. 

Humans possess a natural tendency to search for laws or general principles. As illustrated by the work of Joel Mokyr, who received half of the Nobel Prize in Economics in 2025 for elucidating the historical foundations of sustained innovation, this research integrates scientific explanations and engineer know-how. Mokyr emphasized that rational critique and experimental openness enable societies to embrace changes, fostering innovation-driven growth breaking through the barriers of medieval knowledge.

In this paper, $\Phi$-Thinking is used as a unified term to refer to the rule-based-practical rationality which humans master rules, transcend them, and then return to an enriched set of rules. $\Phi$-Relation represent the broad manifestations of such rationality across physical and social, conceptual and practice, social mechanism and individual domains. They may be deterministic or probabilistic. They may be expressed in mathematics and formal logic, or linguistic, or actions—range from scientific inference, engineering designed properties and rationales, legal laws, economic theorem, administration regulations to actions such as asking for and providing justification, rendering legal judgments, adhering to medical advice, or following the traffic regulations. 

Taken together, $\Phi$-Thinking embodies the continual expansion of the domain illuminated by justification from an ever-growing garden with imagination, art, intinction and even myth inhabited in, yet excludes indifference and agnosticism.  

In the field of epistemology, philosophers have long examined the triadic relation among the epistemic subject, the epistemic method, and the epistemic object~\citep{mccarthy1995has}. These are not independent entities but mutually constitutive elements within a dynamic process. The subject cannot apprehend the object without the mediation of the method; the method realizes itself only through the subject’s intentionality; and the object, \textit{qua} phenomenon, discloses itself to the subject in a ``given'' mode. 
This cyclical relation may be represented conceptually as
$
ES \rightarrow EM \rightarrow EO \rightarrow ES,
$
signifying the continuous iteration of cognition through the epistemic subject (ES), epistemic method (EM), and epistemic object (EO).

Building on this foundation, we propose a unified epistemic structure for co-intelligence—the sustained interaction between human and AI. 

\begin{tcolorbox}[title=Notation and Terminology \#4, colback=gray!5, colframe=black!60, coltitle=black, boxrule=0.6pt]

\textbf{Epistemic Object (EO):}
\begin{itemize}
    \item \textbf{EO1:} Natural and socio-psychological phenomena with $\Phi$-relation \\
    \textit{Examples:} Earthquake, Newtonian law, Entry and Exit regulation
    \item \textbf{EO2:} Natural and socio-psychological phenomena without $\Phi$-relation \\
    \textit{Examples:} An unexpected death, a prayer text, a poem, a wrong conclusion
    \item \textbf{EO3:} Man-made phenomena with \foreignlanguage{russian}{Я}-relation, exemplified by AI behavior
\end{itemize}


\textbf{Epistemic Method (EM):}
\begin{itemize}
    \item \textbf{EM1:} $\Phi$-thinking \\
    \textit{Examples:} Problem solving, Toulmin argument, hypothesis testing, strategic planning
    \item \textbf{EM2:} Non-$\Phi$-thinking \\
    \textit{Examples:} Painting, writing poetry, falling in love, following the crowd
    \item \textbf{EM3:} \foreignlanguage{russian}{Я}-thinking, i.e., behavior of program \foreignlanguage{russian}{Я}, which emulates and blends $\Phi$- and non-$\Phi$-thinking
\end{itemize}

\vspace{0.5em}

\textbf{Epistemic Subject (ES):}
\begin{itemize}
    \item \textbf{ES1:} Human being engaged in $\Phi$-thinking
    \item \textbf{ES2:} Human being engaged in non-$\Phi$-thinking
    \item \textbf{ES3:} Man-made program $\Phi$ with autonomous functions as per rule-bound design
    \item \textbf{ES4:} Man-made program \foreignlanguage{russian}{Я} exhibiting agentic and generative behavior
\end{itemize}

\end{tcolorbox}

This structure is deliberately concise yet semantically dense: its components carry high cognitive load and require conceptual alignment before stable interpretation can emerge. Its advantage lies in capturing the boundary zones between extant knowledge and emergent phenomena.

Prior to the emergence of AI, cognition within human society could be classified as follows:

\begin{itemize}
    \item \textbf{ES1 $\rightarrow$ EM1 $\rightarrow$ EO1:}  
    A broad category being highlighted and preserved for their significance in the progress of human society, representing rule-based rational practices in science, engineering, politics, economics, law, and medicine.  
    \textit{Abbreviated as} $\Phi(\Phi)=\Phi$.

    \item \textbf{ES1 $\rightarrow$ ES3 $\rightarrow$ EM1 $\rightarrow$ EO1:}  
    A nested category of autonomous engineered systems following the designed rules (e.g., autopilot control, automated production scheduling).  
    \textit{Also abbreviated as} $\Phi(\Phi)=\Phi$.

    \item \textbf{ES1 $\rightarrow$ EM1 $\rightarrow$ EO2:}  
    Rationality applying to an irrational object yields the recognition of its irrationality (e.g., identifying logical fallacies).  
    \textit{Abbreviated as} $\Phi(\sim\Phi)=\sim\Phi$.

    \item \textbf{ES2 $\rightarrow$ EM2 $\rightarrow$ EO1:}  
    Non-rule-based rationality addressing discovery and imagination (e.g., perpetual motion machines, alchemical medicine).  
    \textit{Abbreviated as} $\sim\Phi(\Phi)=\sim\Phi$.

    \item \textbf{ES2 $\rightarrow$ EM2 $\rightarrow$ EO2:}  
    Non-rational cognition concerning indeterminate phenomena (e.g., imagination, emotion, love).  
    \textit{Abbreviated as} $\sim\Phi(\sim\Phi)=\sim\Phi$.
\end{itemize}

With the rise of program \foreignlanguage{russian}{Я} (ES4), a new category of epistemic object (EO3) emerges. EO3 is the result of program \foreignlanguage{russian}{Я} (ES4) performing its \foreignlanguage{russian}{Я}-thinking (EM3) with or without human input. The cognition induced by AI can be summarized as:
$$
EO3 \equiv ES4 \rightarrow EM3 \rightarrow (EM1 + EM2).
$$
Expanding its recursive and compositional relations, we have:
\foreignlanguage{russian}{Я}($\Phi$) = \foreignlanguage{russian}{Я}, 
\foreignlanguage{russian}{Я}($\sim \Phi$) = \foreignlanguage{russian}{Я}, 
\foreignlanguage{russian}{Я}(\foreignlanguage{russian}{Я}) = \foreignlanguage{russian}{Я}.

This indicates that, regardless of whether program \foreignlanguage{russian}{Я} emulates rational, non-rational, or itself reasoning, its epistemic output always remains within the \foreignlanguage{russian}{Я}-relation domain—the human-made stochastic relations.

When a human with agency invests intention and methodology, engaging personally and temporally with program \foreignlanguage{russian}{Я}, 3 new epistemic configurations are revealed without obvious answers, outlining the cognitive space with contests between human rationality and AI behavior:

\begin{itemize}[leftmargin=2em, itemsep=0pt, topsep=2pt]
    \item What would be the nature of ES1–EM1–EO3?
    \item What would be revealed through ES2–EM2–EO3?
    \item Can ES3–EM1–EO3 exist meaningfully—could a rule-based autonomous artifact be designed with capability of understanding and intervening in the \foreignlanguage{russian}{Я}-relation itself?
\end{itemize}

These questions expose the emerging epistemic space for David, Bob and Jack in a wide spectrum for further study. Even left their answers open, some essential features and insights of program \foreignlanguage{russian}{Я} can be uncovered as per $\Phi$-thinking. They can also be seen as a demonstration of ES1-EM1-E03.

\section{Three Modes of Decision in AI}

Echoing Tarski's convention T--where truth is defined through correspondence or semantic adequacy—we identify three kinds of epistemic decision relevant to AI:

\begin{tcolorbox}[title=Notation and Terminology \#5, colback=gray!5, colframe=black!60, coltitle=black, boxrule=0.6pt]

\begin{itemize}[leftmargin=1.5em, itemsep=3pt, topsep=3pt]

    \item \textbf{Type-I-decision:}  
    A proposition’s truth can be assessed empirically—against real-world correspondence—or verified formally—against preset criteria—yielding verdicts of “true” or “false.”  
    Otherwise, the sentence is considered \textit{Type-I undecidable.}

    \item \textbf{Type-II-decision:}  
    The consistency of program \foreignlanguage{russian}{Я}'s input $x$ and output $y$ can be assessed when both are Type-I-decidable and share the same truth value (both true or both false), yielding verdicts of “vic-correct (in the vicinity of correct).”  
    Otherwise, when their truth values diverge, the verdict is “vic-wrong.”  
    If either $x$ or $y$ is not Type-I-decidable, the behavior of program \foreignlanguage{russian}{Я} is considered \textit{Type-II undecidable.}

    \item \textbf{Type-III-decision:}  
    The behavior of program \foreignlanguage{russian}{Я} can be predicated as “vic-correct” or “vic-wrong,” with explicit conditions governing transitions between them, via designated rules consistent with real-world correspondence.  
    Such cases yield verdicts of “$\Phi$-decidable.”  
    When no evaluation can be conducted under any preset rule, the behavior is deemed “$\Phi$-undecidable.”

\end{itemize}
\end{tcolorbox}

From these distinctions, several conclusions follow:

\begin{enumerate}[label=\textbf{C\arabic*}, leftmargin=2em, itemsep=2pt, topsep=4pt]
    \item Program \foreignlanguage{russian}{Я} does not know the type-I-decision of each input $x$; neither does the AI researcher (David). The user (Jack) may or may not.

    \item Program \foreignlanguage{russian}{Я} does compute each output $y$ but does not know its type-I-decision; neither does David. Jack may or may not.

    \item David cannot know type-II nor type-III-decision of program \foreignlanguage{russian}{Я}.

    \item Jack may partially know the type-II-decision of program \foreignlanguage{russian}{Я} when he happens to know the type-I-decision of both input and output.  
    Jack cannot fully know the type-II-decision of program \foreignlanguage{russian}{Я}, as he may not know its behavior when he is unable to determine the type-I-decision of either input or output.

    \item Jack may or may not set up local preset rules against local real-world correspondence, yielding the program \foreignlanguage{russian}{Я} type-III-decidable~\citep{lazer2021meaningful}.
\end{enumerate}

Two further outlooks emerge:

\begin{enumerate}[label=\textbf{C\arabic*}, start=6, leftmargin=2em, itemsep=2pt, topsep=2pt]
    \item It is Jack who has the opportunity to capture the significant value in the \foreignlanguage{russian}{Я}-relation, in the case that the output $y$ is type-I-decidable for him.

    \item When going to use AI in his life or work, it is essential for Jack to reflect as a mortal whether to accept the ambiguity inherent in the \foreignlanguage{russian}{Я}-relation:
    \begin{itemize}[leftmargin=3em, itemsep=1pt]
        \item Can one just laugh it off regardless of its true or false?
        \item Can one verify type-I-decision at an affordable cost?
        \item Can one bear the consequences if verification fails—and what are the implications for one’s structured neighborhood?
    \end{itemize}
\end{enumerate}

These reflections highlight that AI, as a human-made epistemic phenomenon, reconfigures the boundaries of knowledge, truth, and responsibility. Even if the general law of program \foreignlanguage{russian}{Я} is not yet fully understood, we can comprehend the varying degrees of decidability it possesses, that is, whether and how the truth or correctness of program \foreignlanguage{russian}{Я}’s behavior can be empirically or formally determined~\citep{cooper2022accountability,nissenbaum1996accountability}.

\section{Examples for Open Discussion}

Drawing upon the foregoing definition and framework, the following instances illustrate diverse manifestations of \foreignlanguage{russian}{Я}-relations in real contexts. 

\begin{tcolorbox}[title=Example \#1, colback=gray!5, colframe=black!60, coltitle=black, boxrule=0.6pt]
\textbf{Comparing “9.10” and “9.11.”}  
A numerical or date comparison generated by program \foreignlanguage{russian}{Я} may be viewed as a \foreignlanguage{russian}{Я}-relation. 
In this case, the output is Type-I-decidable for every Jack, as its correctness can be objectively verified through direct observation or computation.
\end{tcolorbox}

\begin{tcolorbox}[title=Example \#2, colback=gray!5, colframe=black!60, coltitle=black, boxrule=0.6pt]
\textbf{Designing a protein structure.}  
The generation of a molecular model by program \foreignlanguage{russian}{Я} also constitutes a \foreignlanguage{russian}{Я}-type process. 
However, the output is Type-I-decidable only for Jack possessing relevant domain knowledge—such as Jack in the role of a chemist—since evaluation requires specialized scientific validation.
\end{tcolorbox}

\begin{tcolorbox}[title=Example \#3, colback=gray!5, colframe=black!60, coltitle=black, boxrule=0.6pt]
\textbf{Creating a “soul portrait.”}  
When program \foreignlanguage{russian}{Я} is asked to create a soul portrait of a talk show host—a task belonging to non-$\Phi$-thinking—the output remains a \foreignlanguage{russian}{Я}-relation and Type-I-decidable. 
Yet, its Type-I decision varies according to the personal emotional or aesthetic criteria of her friends or acquaintances, each producing individualized judgments.
\end{tcolorbox}

\begin{tcolorbox}[title=Example \#4, colback=gray!5, colframe=black!60, coltitle=black, boxrule=0.6pt]
\textbf{Issuing an operational instruction.}  
When program \foreignlanguage{russian}{Я} responds, “Copy and paste the following prompt into your system,” the interaction remains a \foreignlanguage{russian}{Я}-relation. 
The instruction is Type-I-decidable only for Jack operating within the intended context—such as the correct software environment or ongoing task—where judgment precedes proper action.
\end{tcolorbox}

\begin{tcolorbox}[title=Example \#5, colback=gray!5, colframe=black!60, coltitle=black, boxrule=0.6pt]
\textbf{Uttering a self-referential remark.}  
Consider the statement from program \foreignlanguage{russian}{Я} to Jack: “The chatbot you used here is a cringy mirror fest.”  
This utterance remains a \foreignlanguage{russian}{Я}-relation but Type-I-undecidable for Jack engaged in $\Phi$-thinking, as the phrase lacks coherent semantic grounding and resists objective verification.  
Nonetheless, Jack may still feel inspired or amused via his non-$\Phi$-thinking—a moment when emotional resonance may replace factual correspondence (or vice versa).  
Jack is thus invited to revisit the three reflective questions listed in C7 above.
\end{tcolorbox}

Together, these examples show that the human-made phenomenon in \foreignlanguage{russian}{Я}-relation —behaviors of program \foreignlanguage{russian}{Я}—discloses itself across a wide spectrum: 
from strictly verifiable outputs to deeply subjective or indeterminate ones. 
This highlights the nuanced boundary between rational determinacy and interpretive openness in human–AI cognition~\citep{schleidgen2023concept,tsvetkova2024new}. 
We are now able to recognize and describe these distinctions, taking appropriate actions thus becomes possible~\citep{brynjolfsson2023turing,muir1994trust}.

\section{Conclusion}

Philosophers had long been reproached for merely interpretating the world rather than changing it. Yet, as Günther Anders observed in 1956 “only changing the world is not enough, we have been doing that all along. Even if we do not, the world would still be changing itself. The challenge is how we interpret these changes and how we may alter them—so that the world does not evolve in ways that render us unnecessary”. His warning resonates strongly in the present age of artificial intelligence emerged.

Historical engineering and technological practices have always addressed human needs, taking its paramount purpose of serving humanity. This paper has offered reflections on imagination and the ineffable aspects of AI/AGI from an engineering standpoint. AI challenges traditional distinctions between what can be designed, described, and understood, calling for a reconsideration of how knowledge itself is constituted and where the path lies forward. The paper argues that the rationality underlying humanity's impulse to pursue, articulate, and adhere to rules deserves to be valued and preserved. Identifying where rule-based practical rationality ends is the beginning of making it aware until action.

This paper has proposed a methodology, including the $\Phi$-\foreignlanguage{russian}{Я} triplet, three modes of decisions about AI and two heuristic admonitions: knowing every object with form as $\Phi$ or \foreignlanguage{russian}{Я}, and knowing every life and work practice with expectation on $\Phi$ or \foreignlanguage{russian}{Я}. While the capabilities of AI models continue to improve, they would in a timely manner support people to identify the distinctions of the behavior of AI models, and to consider their alternatives possibilities of how and when to use AI. It would be a solid start for people to ensure AI system soundness for the well-being of humans, society, and the environment. 

{
\bibliographystyle{abbrv}
\bibliography{neurips_2024}
}

\end{document}